# Computational lattice-gas modeling of the electrosorption of small molecules and ions

P.A. Rikvold [a,b,1,2], M. Gamboa-Aldeco [a,c,3], J. Zhang [a],
M. Han [a,c,4], Q. Wang [a,b] H.L. Richards [a,b], and A. Wieckowski [c,2]

[a] *Supercomputer Computations Research Institute, Florida State University,
Tallahassee, Florida 32306-4052, USA*

[b] *Center for Materials Research and Technology, and Department of Physics,
Florida State University, Tallahassee, Florida 32306-3016, USA*

[c] *Department of Chemistry and Frederick Seitz Materials Research Laboratory,
University of Illinois, Urbana, Illinois 61801, USA*

**Abstract**

We present two recent applications of lattice-gas modeling techniques to electrochemical adsorption on catalytically active metal substrates: urea on Pt(100) and (bi)sulfate on Rh(111). Both systems involve the specific adsorption of small molecules or ions on well-characterized single-crystal electrodes, and they provide a particularly good fit between the adsorbate geometry and the substrate structure. The close geometric fit facilitates the formation of ordered submonolayer adsorbate phases in a range of electrode potential positive of the range in which an adsorbed monolayer of hydrogen is stable. In both systems the ordered-phase region is separated from the adsorbed-hydrogen region by a phase transition, signified in cyclic voltammograms by a sharp current peak. Based on data from *in situ* radiochemical surface concentration measurements, cyclic voltammetry, and scanning tunneling microscopy, and *ex situ* Auger electron spectroscopy and low-energy electron diffraction, we have developed specific lattice-gas models for the two systems. These models were studied by group-theoretical ground-state calculations and numerical Monte Carlo simulations, and effective lattice-gas interaction parameters were determined so as to provide agreement with the experimental results.

[1] Electronic address: rikvold@scri.fsu.edu.

[2] Authors to whom correspondence should be addressed.

[3] Permanent address: Lynntech, Inc., 7610 Eastmark Dr., Suite 105, College Station, TX 77840, USA.

[4] Permanent address: Department of Physics, Seoul City University, Seoul 130-743, Republic of Korea.

# 1 Introduction

Both single-component and multi-component adsorption processes are strongly influenced by lateral adsorbate-adsorbate interactions. A theoretical approach to the investigation of these effects is the formulation of lattice-gas models and their solution by analytical and numerical techniques. This method has previously been successfully applied to phase transitions at solid-vacuum and solid-gas interfaces [1]. Provided care is taken to consider its limitations, the lattice-gas modeling approach is also useful to understand the statistical-mechanical aspects of multicomponent adsorption at solid-liquid interfaces, in particular the complicated adsorbate phase diagrams observed near the potential of zero charge [2–13].

In this paper we discuss the application of computational lattice-gas modeling to two different experimental systems involving adsorption on well-characterized single-crystal electrodes: urea on Pt(100) [14–19] and (bi)sulfate on Rh(111) [20–22]. Both systems exhibit a dramatic peak sharpening in the cyclic voltammogram (CV) from several hundred mV to on the order of 10 mV when the adsorbate species (urea or (bi)sulfate) is added to the supporting perchloric-acid electrolyte [14,20]. We associate this effect with a phase transition in the layer of contact adsorbed particles. This transition involves the replacement of a monolayer of adsorbed hydrogen on the negative-potential side of the CV peak by an ordered submonolayer of urea or (bi)sulfate, respectively, on the positive-potential side.

The peak-sharpening phenomenon is much weaker or absent when the same substances are electrosorbed onto other crystal planes of the same metals [14,20]. This high specificity with respect to the adsorbent surface structure indicates that the effect depends crucially on the geometric fit between (at least one of) the adsorbate species and the surface. Guided by this observation, we have developed specific lattice-gas models with three adsorbed species: water, hydrogen, and either urea or (bi)sulfate, depending on the system. These models were studied by group-theoretical ground-state calculations [23–25] and numerical Monte Carlo (MC) simulations [26]. Based on the results of these calculations, we adjusted the effective lateral adsorbate-adsorbate interactions in the models to fit experimentally measured quantities. In addition to the systems discussed here, similar modeling strategies, differing mainly in the means by which thermodynamic information was extracted from the lattice-gas Hamiltonian, have been applied to other systems involving localized adsorption of two species of small molecules or ions [3–12]. For the systems discussed here, the experiments with which agreement was sought include cyclic voltammetry, *in situ* radiochemical surface concentration measurements (RCM), and *ex situ* Auger electron spectroscopy (AES) and low-energy electron diffraction (LEED).

Our lattice-gas models are defined through a generalization of the standard three-state lattice-gas Hamiltonian (energy function) used, *e.g.*, in Refs. [3–6]:

$$\mathcal{H}_{\mathrm{LG}} = \sum_n \left[ -\Phi_{\mathrm{AA}}^{(n)} \sum_{\langle ij \rangle}^{(n)} c_i^{\mathrm{A}} c_j^{\mathrm{A}} - \Phi_{\mathrm{AH}}^{(n)} \sum_{\langle ij \rangle}^{(n)} \left( c_i^{\mathrm{A}} c_j^{\mathrm{H}} + c_i^{\mathrm{H}} c_j^{\mathrm{A}} \right) - \Phi_{\mathrm{HH}}^{(n)} \sum_{\langle ij \rangle}^{(n)} c_i^{\mathrm{H}} c_j^{\mathrm{H}} \right] + \mathcal{H}_3 - \bar{\mu}_{\mathrm{A}} \sum_i c_i^{\mathrm{A}} - \bar{\mu}_{\mathrm{H}} \sum_i c_i^{\mathrm{H}} \,. \tag{1}$$

Here $c_i^{\mathrm{X}} \in \{0,1\}$ is the local occupation variable for species X [X=A (adsorbate) or H (hydrogen)], and the third adsorption state ("empty" or "solvated") corresponds to $c_i^{\mathrm{A}} = c_i^{\mathrm{H}} = 0$. The sums $\sum_{\langle ij \rangle}^{(n)}$ and $\sum_i$ run over all $n$th-neighbor bonds and over all adsorption sites, respectively, $\Phi_{\mathrm{XY}}^{(n)}$ denotes the effective XY pair interaction through an $n$th-neighbor bond, and $\sum_n$ runs over the interaction ranges. The term $\mathcal{H}_3$ contains three-particle [27] and possibly multi-particle interactions. Both the interaction ranges and the absence or presence of multi-particle interactions depend on the specific system. The change in electrochemical potential when one X particle is removed from the bulk solution and adsorbed on the surface is $-\bar{\mu}_{\mathrm{X}}$. The total number of surface unit cells is $N$, and $\Theta_{\mathrm{X}} = N^{-1} \sum_i c_i^{\mathrm{X}}$ is the surface coverage by species X. The sign convention is such that $\Phi_{\mathrm{XY}}^{(n)} > 0$ denotes an effective attraction, and $\bar{\mu}_{\mathrm{X}} > 0$ denotes a tendency for adsorption in the absence of lateral interactions. Straightforward modifications of the above form are necessary if the adsorption sites for the two species are different, as they are in the two cases studied here. We emphasize that the interactions in Eq. (1) are *effective* interactions mediated through a variety of channels, including electrons, surface phonons, and the fluid [5], and that they therefore in principle may depend on temperature and electrode potential.

The electrochemical potentials in Eq. (1) are (in the weak-solution approximation) related to the bulk concentrations [X] and the electrode potential $E$ as

$$\bar{\mu}_{\mathrm{X}} = \bar{\mu}_{\mathrm{X}}^0 + RT \ln \frac{[\mathrm{X}]}{[\mathrm{X}]^0} - z_{\mathrm{X}} F E \,, \tag{2}$$

where $R$ is the molar gas constant, $T$ is the absolute temperature, $F$ is Faraday's constant, and the effective electrovalence of X is $z_{\mathrm{X}}$. The quantities superscripted with a 0 are reference values which contain the local binding energies to the surface.

For a given set of effective interactions, the coverages $\Theta_{\mathrm{X}}$ as functions of the electrochemical potentials and temperature are calculated numerically by MC simulation. In the absence of diffusion and double-layer effects and in the limit that the potential sweep rate $dE/dt \to 0$, the CV current per unit cell of the surface (oxidation currents considered positive) is related to the lattice-gas response functions $\partial \Theta_X / \partial \bar{\mu}_Y$ as [16–18]

$$i = eF \left\{ z_A^2 \frac{\partial \Theta_A}{\partial \bar{\mu}_A} + 2z_A z_H \frac{\partial \Theta_H}{\partial \bar{\mu}_A} + z_H^2 \frac{\partial \Theta_H}{\partial \bar{\mu}_H} \right\} \frac{dE}{dt} , \qquad (3)$$

where $e$ is the elementary charge unit.

Once a lattice-gas Hamiltonian has been specified, the corresponding equilibrium thermodynamic properties can be studied by a number of analytical and numerical methods, depending on the quantities of interest and the complexity of the Hamiltonian. In addition to group-theoretical ground-state calculations to determine the domains of stability of the various possible phases [23–25], finite-temperature properties have been obtained by mean-field approximations [7] (although these can be unreliable for low-dimensional systems with short-range interactions [28,29]), Padé-approximant methods based on liquid theory [9–12], numerical transfer-matrix calculations [3–6], and Monte Carlo simulations [4,16–18].

## 2  Urea on Pt(100)

In addition to the aforementioned, surface-specific narrowing of the CV peak upon the addition of urea to the supporting electrolyte, the experimental observations to which we have attempted to fit the model are as follows. (For experimental details, see Ref. [18].)

1. The urea coverage $\Theta_U$, measured *in situ* by RCM, changes over a potential range of approximately 20 mV around the CV peak position from near zero on the negative side to approximately 1/4 molecules per Pt(100) unit cell (1/4 monolayers, or ML) on the positive side (0.26±0.04 ML at −60 mV versus Ag/AgCl), as shown in Fig. 1b. Corresponding CV curves are shown in Fig. 1a.

2. *Ex situ* AES studies of the electrode emersed at −60 mV are consistent with the RCM result ($\Theta_U$=0.24±0.03 ML).

3. *Ex situ* LEED studies of the electrode emersed at −60 mV show an ordered c(2×4) adsorbate structure, consistent with an ideal coverage of 1/4 ML. Upon emersion on the negative side of the CV peak, only an unreconstructed (1×1) surface (presumably prevented from reconstructing by trace impurities) was found.

4. The CV peak potentials were measured for [U] between 0.3 and 3.0 mM. In Fig. 2a are shown the corresponding electrochemical potentials (× and dotted line), obtained from Eq. (2) with the same parameters used to fit Fig. 1.

5. In addition to the above experiments, which were all performed at room temperature, cyclic voltammetry was performed at several fixed temperatures

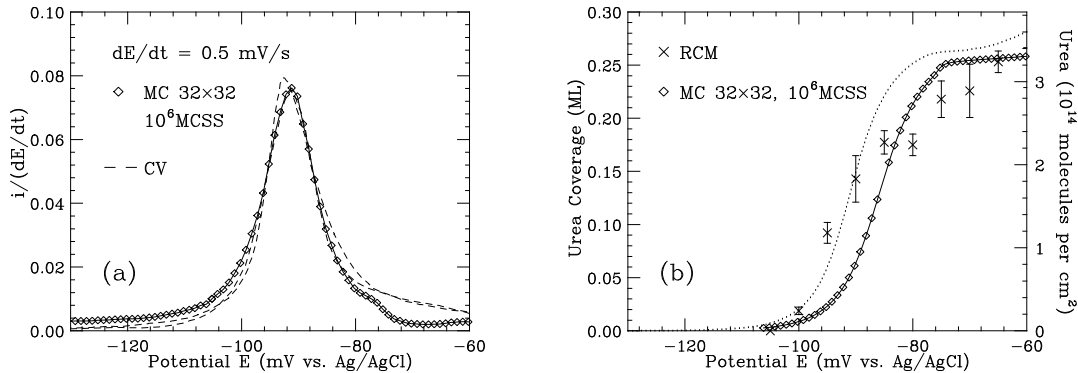

Fig. 1. Urea on Pt(100) in 0.1 M HClO$_4$ at room temperature. (a) Experimental (dashed curves) and simulated ($\diamond$ and solid curve) normalized CV currents, $i/(\mathrm{d}E/\mathrm{d}t)$ in elementary charges per mV per Pt(100) unit cell, at 1.0 mM bulk urea. The two dashed curves are representative negative-going voltammograms, and their differences indicate the experimental uncertainty. (b) Urea coverage measured by RCM at 0.5 mM bulk urea ($\times$ with error bars) and simulated by MC ($\diamond$ and solid curve), shown versus electrode potential. Also shown as a dotted curve is the simulation result for 1.0 mM bulk urea, which corresponds directly to the simulated CV profile shown in (a). The effective interactions used in this work are (in kJ/mol) $\Phi_{HH}^{(1)}=-2.0$, $\Phi_{HU}^{(1)}=-8.0$, $\Phi_{HU}^{(2)}=-4.0$, $\Phi_{UU}^{(1)}=-13.0$, $\Phi_{UU}^{(2)}=-10.0$, $\Phi_{UU}^{(3)}=-5.9$, $\Phi_{UU}^{(4)}=-0.5$, $\Phi_{UU}^{(5)}=-2.5$, $\Phi_{UU}^{(6)}=-3.0$, $\Phi_{UU}^{(7)}=+0.25$, $\Phi_{UU}^{(8)}=-2.0$, the effective electrovalences were taken as $z_H=+1$ and $z_U=-1$, and the room-temperature simulations were performed at 28°C. The results shown are for a system of 32×32 Pt(100) unit cells. The statistical uncertainty in the MC results is everywhere much smaller than the symbol size.

between 0°C and 40°C.

The lattice-gas model developed to account for these observations is based on the assumption that urea [CO(NH$_2$)$_2$] coordinates the platinum surface through its nitrogen atoms (or NH$_2$ groups), with the C=O group pointing away from the surface. Since the unstrained N-N distance in urea matches the lattice constant of the square Pt(100) surface quite well (2.33 Å [30] versus 2.77 Å [31]), we assume that urea occupies two adsorption sites on the square Pt(100) lattice. Integration of the CV curves indicates that the hydrogen saturation coverage in the negative-potential region corresponds to one elementary charge per Pt(100) unit cell, and that most of the surface hydrogen is desorbed in the same potential range where urea becomes adsorbed. Therefore, we assumed [15] that hydrogen adsorbs in the same on-top positions as the urea nitrogen atoms. This assumption was recently confirmed by visible-infrared sum generation spectroscopy [32]. The resulting model [15] is a dimer-monomer model in which hydrogen is adsorbed at the nodes and urea on the bonds of a square lattice representing the Pt(100) surface. Simultaneous occupation by urea of two bonds that share a node is excluded, as is occupation by hydrogen of a node adjacent to a bond occupied by urea. In order to stabilize the observed c(2×4) phase, effective interactions are included through eighth-nearest neighbors [16–19], as shown in Fig. 5 of Ref. [18]. The model is illustrated in Fig. 3.

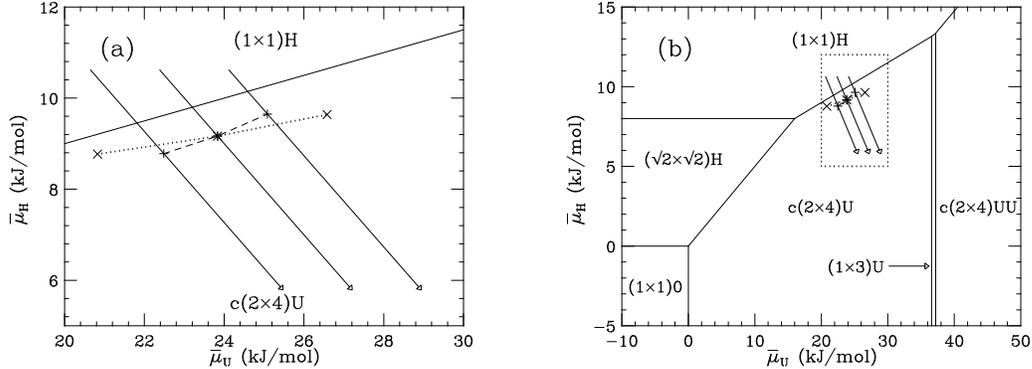

Fig. 2. Phase diagrams for urea on Pt(100) in 0.1 M HClO$_4$, shown in the $(\bar{\mu}_\mathrm{U}, \bar{\mu}_\mathrm{H})$ plane. (a) The zero-temperature phase boundary between the c(2×4) phase with $\Theta_\mathrm{U}=1/4$, indicated as c(2×4)U in the figure, and the (1×1) phase with $\Theta_\mathrm{H}=1$, indicated as (1×1)H, is shown as a solid line together with the electrochemical potentials corresponding to experimental (× connected by dotted lines) and simulated (+ connected by dashed lines) CV peak positions. All simulation parameters are as in Fig. 1. The solid arrows represent positive-going $E$ scans from $-106$ mV to $-56$ mV. From left to right they represent [U]=2 mM, 1 mM, and 0.5 mM. The experimental peak positions are averages over positive-going and negative-going scans with $|\mathrm{d}E/\mathrm{d}t| = 50$ mV/s, but the average positions depend little on $|\mathrm{d}E/\mathrm{d}t|$. (b) A full ground-state diagram for the interactions used, showing all the phases present. The phase indicated in the figure by c(2×4)UU has $\Theta_\mathrm{U}=1/2$, (1×3)U has $\Theta_\mathrm{U}=1/3$, $(\sqrt{2}\times\sqrt{2})$H has $\Theta_\mathrm{H}=1/2$, and (1×1)0 is the empty lattice. The phase regions outside the dotted box, which corresponds to panel (a), are likely not experimentally accessible.

The effective lattice-gas interactions (given in the caption of Fig. 1) were determined by a ground-state calculation followed by numerical MC simulation. Since the number of model parameters is large, the numerical calculations are time consuming, and the experimental data concern a number of different quantities, parameter estimation by a formal optimization procedure was not a practical alternative. (This contrasts with the simpler situations discussed in Refs. [5,29,33], where a small number of lattice-gas parameters could be determined by a formal least-squares procedure to fit extensive experimental results for a single thermodynamic quantity.) The model parameters were therefore varied "by hand", taking into consideration both the various experimental results and available chemical and physical background information, until acceptable agreement was obtained with room-temperature experimental results. In particular, we sought agreement between the simulated CV profiles and those obtained experimentally at the lowest practical potential scan rate of 0.5 mV/s, while at the same time ensuring that the simulated ordered phase on the positive side of the CV peak corresponded to the observed c(2×4) structure with 1/4 ML urea coverage. Whereas this procedure may be criticized as unsystematic, we believe it makes the maximum use of all the available experimental and theoretical information. A detailed discussion of the roles of the different effective interactions is given in Ref. [18].

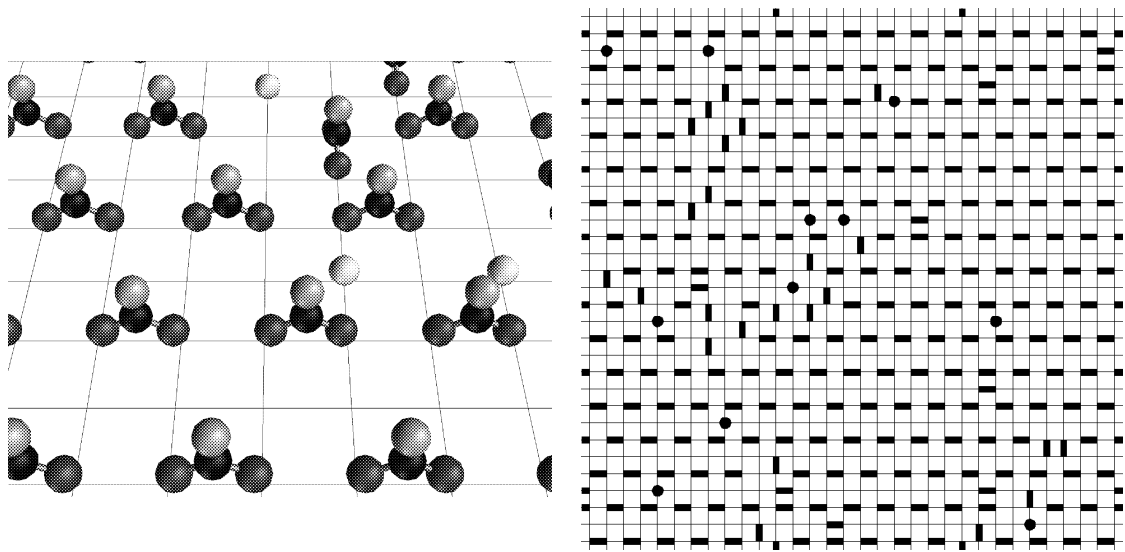

Fig. 3. Left panel: Illustration of the microscopic model, showing urea molecules with their NH$_2$ groups (dark gray) on the square Pt(100) lattice sites and their C=O groups (C black and O lighter gray) pointing away from the surface. The hydrogen atoms are shown as single light gray spheres. Right panel: A typical equilibrium configuration generated by MC in the ordered c(2×4) phase region at −60 mV. Urea molecules are shown as black rectangles on the lattice bonds and hydrogen as black circles at the nodes. Locally disordered fluctuations in the ordered background are seen, as well as a low density of hydrogen on vacant sites between the urea molecules. This figure is adapted from the video, Ref. [19].

The numerical simulations, which used systems with up to 32×32 square-lattice unit cells, were performed on a cluster of IBM RS6000 workstations. We used a heat-bath MC algorithm [26] with updates of clusters consisting of five nearest-neighbor nodes arranged in a cross, plus their four connecting bonds. After symmetry reductions these clusters have 64 different configurations, and the corresponding code is rather slow in terms of machine time per MC step. However, the additional transitions allowed by these clusters, relative to minimal clusters consisting of two sites and their connecting bond, include "diffusion-like" moves in which the urea molecules can go from one bond to another and the hydrogen atoms from one site to another, without changing the local coverages within the cluster. These moves significantly reduce the free-energy barriers that must be surmounted in order to locally minimize the adsorbate energy, and consequently dramatically reduce the number of MC steps per site (MCSS) necessary for the system to reach thermodynamic equilibrium. This speedup is particularly evident deep in the c(2×4) phase, where an algorithm based on updates of minimal clusters proved unable to equili-brate the system, even after several million MCSS. Thermodynamic quantities were calculated as averages over each MC run, with samples taken every fifty MCSS.

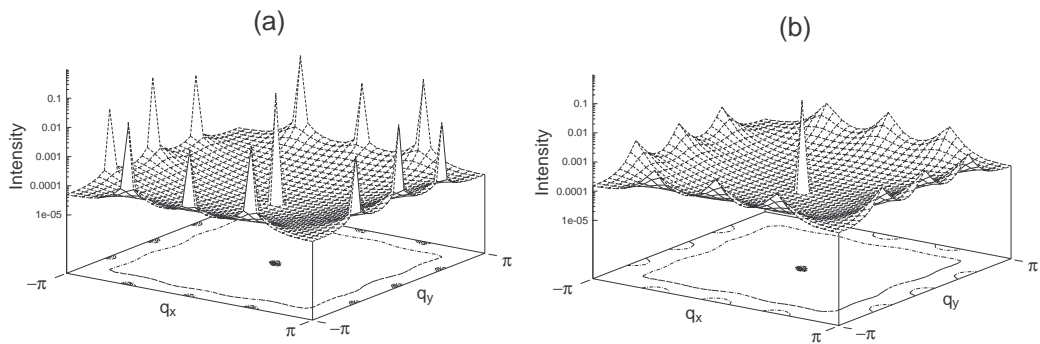

Fig. 4. Simulated "LEED patterns" in the Brillouin zone surrounding a Bragg peak. The simulated scattering intensity is shown on a logarithmic scale versus the $x$- and $y$-components of the dimensionless scattering vector, $q_x$ and $q_y$. Contour plots of the intensity are shown in the $(q_x, q_y)$ plane. Both panels are for 1 mM urea. (a): $E = -0.72$ mV. (b): $E = -0.81$ mV. See discussion in the text.

Near the CV peak at $-91$ mV, the simulated height and width of which agree well with the room-temperature experimental results as shown in Fig. 1a, no finite-size effects were observed for system sizes above 8×8. (For a discussion of temperature effects, see below.) This indicates that this peak in the model is not directly due to a phase transition [34], but probably results from strong, local coverage fluctuations in a disordered phase near a tricritical point. At the slightly more positive potential of $-75$ mV a second-order, order-disorder transition (probably in the universality class of the $XY$ model with cubic anisotropy) occurs, which is indicated by a sharp peak in the response function for the c(2×4) order parameter. That this peak indeed signifies a phase transition, is indicated by its finite-size scaling behavior [34]. Since it does not involve large changes in the coverages, in the room-temperature simulation this order-disorder transition only shows up in the CV as a small shoulder on the positive side of the peak.

Simulated "LEED patterns" were obtained as the squared Fourier transform of the adsorbed urea configurations. These were obtained by the Fast Fourier Transform algorithm (our program was based on the subroutine `fourn` of Ref. [35]) and averaged in the same way as the thermodynamic quantities. In Fig. 4 we show results for 1 mM urea at $-72$ mV, which is in the c(2×4) region on the positive side of the order-disorder transition (Fig. 4a), and at $-81$ mV, which lies between the order-disorder transition and the CV peak (Fig. 4b). The triple intensity maxima at the edges of the Brillouin zone, which are characteristic of c(2×4) order, are clearly seen. In the ordered phase they are comparable in intensity with the peak at the zone center, the intensity of which is the square of the urea coverage. In contrast, the c(2×4) maxima in Fig. 4b are much weaker than the peak at the zone center. This indicates that the long-range order is almost totally absent at this potential, even though the urea coverage is near that of the ordered phase.

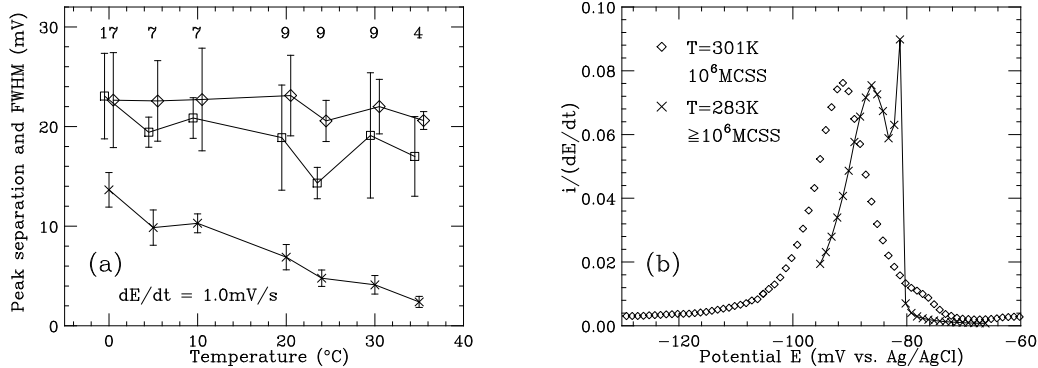

Fig. 5. Temperature effects in experiments and model. (a) The measured temperature dependence of the full width at half maximum of the CV peak (FWHM) during positive-going ($\diamond$) and negative-going ($\square$) scans, and the peak separation ($\times$). For clarity, the FWHM data points are offset slightly to the right ($\diamond$) and left ($\square$) from the actual temperature values. Along the upper edge we give the number of independent scans at each temperature, and the error bars correspond to the empirical standard deviation. (b) Simulated CV profiles for a $32 \times 32$ system at $28°C$ ($\diamond$, identical to Fig. 1a) and $10°C$ ($\times$ and solid line). All model parameters are as in Fig. 1.

Exploratory cyclic voltammetry experiments have been carried out between $0°C$ and $40°C$. The temperature was adjusted by immersing the electrochemical cell in a thermostat controlled water bath. A number of independent potential cycles were carried out at each temperature. In Fig. 5a we show the results for the full width at half maximum (FWHM) and the potential separation between the positive-going and the negative-going peak at $|dE/dt| = 1.0\,\mathrm{mV/s}$. Only temperatures for which at least four potential cycles were performed are included. The data shown are averages over the number of cycles indicated along the upper edge of the figure, and the error bars represent one empirical standard deviation. No statistically significant temperature dependence was found for the FWHM. The same was the case for the maximum currents and the average peak positions, neither of which are shown here. By contrast, the peak separation at fixed $|dE/dt| = 1.0\,\mathrm{mV/s}$ was reduced approximately by a factor of six from the lowest to the highest temperature shown. Whereas this reduction is highly statistically significant, the temperature range is too small to distinguish reliably between, e.g., a linear decrease and an Arrhenius-type behavior on the basis of these experiments alone. However, the reduction with increasing $T$ of the peak separation at constant scan rate indicates the presence of significant temperature-dependent hysteresis, even at this low scan rate.

For comparison with these experimental results, we present in Fig. 5b simulated CV profiles at 10°C and 28°C (the latter identical to the one shown in Fig. 1a). Notice that the shape and height of the non-scaling peak which dominates the simulated CV current at 28°C is hardly changed at all when the temperature is lowered to 10°C. In contrast, the scaling peak corresponding to the order-disorder phase transition, which at 28°C appears only as a small shoulder near the foot of the main peak, at 10°C appears as a sharp peak with an integral corresponding to approximately 15% of the total transported charge.

The simulated CV profile at this temperature does not agree well with the experimentally observed profiles, which do not differ materially from those of the room-temperature voltammograms shown in Fig. 1a. The main difference between the simulated profiles at the two temperatures lies in the sharp spike associated with the order-disorder phase transition. We believe this discrepancy between the simulated and experimental CV profiles at low temperatures can be explained by the temperature-dependent kinetic effects evidenced by the experimental peak separations shown in Fig. 5a. Such effects could in principle be due to bulk diffusion, as well as to surface kinetics including diffusion and ordering. For the bulk concentration range used in this study, bulk diffusion should be unimportant, and we therefore believe the hysteresis is due to processes on the surface. It is difficult to quantify the influence of these kinetic effects on the CV peak shapes and positions without either employing highly phenomenological approximations or carrying out extensive nonequilibrium simulations. Qualitatively, however, one can point out that whereas the characteristic times associated with the wide peak, which reflects only *local* coverage fluctuations on the surface, should be relatively short (estimated from the experimental peak separations as 1–10 s), the timescale associated with the narrow ordering peak should be considerably longer. Two distinct physical phenomena contribute to this effect. The first is the critical slowing-down associated with the continuous order-disorder phase transition [26]. The second is a "gridlock" phenomenon [36] which prevents a disordered urea layer with a coverage near 1/4 ML from easily adjusting to the ordered equilibrium configuration, due to the large free-energy barriers that prevent desorption that could momentarily ease the crowding. This effect is closely related to the dramatic slowing-down suffered in the ordered c(2×4) phase region by the MC algorithms that do not allow lateral diffusive moves. As a result, the narrow peak in the simulated "equilibrium" CV profile should be much more seriously affected by the surface kinetics, to the extent of becoming smeared into an almost undetectable shoulder on the side of the broader peak in the experimental voltammogram. (Recent reviews of related issues of metastability and hysteresis in a variety of fields are found in Refs. [37,38].)

In conclusion, concerning the discussion in the previous paragraph, we believe that the observed kinetic effects are too large to allow us to determine whether or not the sharp ordering peak predicted by the lattice-gas model at the lower temperatures is present in the equilibrium behavior of the experimental system. We can suggest several possible methods by which one could resolve this question. First, voltammetry could be performed at lower scan rates in order to attempt to resolve any sharp ordering peak. However, the required scan rates may well be too low to be practical [39]. Second, one could relegate the CV data to a secondary position in the hierarchy of experimental results to be fitted, concentrating instead on RCM coverage measurements, which are obtained much closer to equilibrium conditions. This is the approach taken in our study of (bi)sulfate adsorption on Rh(111), discussed in Sec. 3. However, the uncertainties in the RCM coverage measurements (5–10%) and electrode potential determination ($\approx \pm 2\,\mathrm{mV}$) makes detection by this method of a narrow peak representing a small total charge very difficult. Third, one could attempt to account for the effects of the surface kinetics on the CV peak shapes and separations, either analytically, as it was recently done in a study of underpotential deposition of copper on Au(111) in sulfuric acid [11], or by kinetic MC simulations. However, the MC algorithms used in the present study were designed to equilibrate the system in a reasonably short amount of computer time, rather than to faithfully reproduce the actual ordering kinetics. We therefore have not attempted to systematically study the influence of the lattice-gas kinetics on the observed voltammograms at finite scan rates. This problem will be considered in the future.

## 3 (Bi)sulfate on Rh(111)

In this section we report some preliminary results of our ongoing study of (bi)sulfate electrosorption on Rh(111) single crystals. Details will be reported elsewhere [21,22]. In addition to the surface-specific narrowing of the CV peak upon the addition of sulfuric acid to the supporting electrolyte, the experimental observations to which we have attempted to fit a lattice-gas model are as follows.
1. The (bi)sulfate coverage $\Theta_S$, measured *in situ* by RCM, changes over a potential range of approximately 15 mV around the CV peak position from near zero on the negative side to approximately 1/5 molecules per Rh(111) unit cell (1/5 ML) on the positive side ($0.21\pm0.01$ ML at $-140$ mV versus Ag/AgCl), as shown in Fig. 6.
2. *In situ* scanning tunneling microscopy (STM) studies of (bi)sulfate on Au(111) [40–42] and Rh(111) [42] indicate an ordered ($\sqrt{3}\times\sqrt{7}$) sulfate submonolayer at potentials sufficiently far to the positive side of the CV peak. The most likely (bi)sulfate coverage deduced from these studies is approximately 1/5 ML, possibly with additional coadsorbed water or hydronium ions [41].

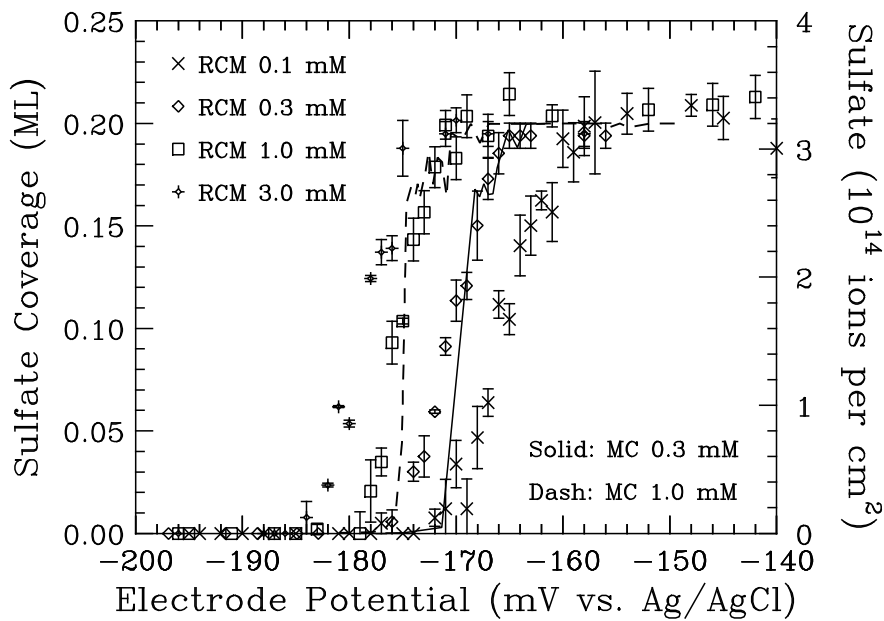

Fig. 6. (Bi)sulfate coverages in 0.1 M HClO$_4$ supporting electrolyte, as measured by the *in situ* RCM technique for bulk sulfuric acid concentrations 0.1 mM ($\times$), 0.3 mM ($\diamond$), 1.0 mM ($\square$), and 3.0 mM ($+$). Together with the experimental data are shown preliminary MC simulation results for 0.3 mM (solid curve) and 1.0 mM (dashed curve), using the simplified lattice-gas model described in the text.

3. *Ex situ* AES studies of the electrode emersed at $-140$ mV are consistent with the RCM result, yielding values of $\Theta_S$ near 1/5 ML [43].
4. *Ex situ* LEED studies of the electrode emersed at $-140$ mV show an ordered ($\sqrt{3}\times\sqrt{3}$) structure [43]. The ideal coverage of this structure (1/3 ML) is not consistent with the *in situ* RCM and STM and the *ex situ* AES results. However, it is possible that as the (bi)sulfate adsorbate system becomes an effectively closed system upon emersion, the change of environment may change the effective interactions (possibly through desorption of strongly coadsorbed water or hydrogen) to cause condensation into ($\sqrt{3}\times\sqrt{3}$) islands coexisting with patches of essentially empty surface, to give a total coverage near 1/5 ML. This explanation has recently been suggested for analogous observations for (bi)sulfate on Au(111) [44].
5. The CV peak potentials were measured for bulk sulfuric acid concentrations between 0.1 and 10 mM.

The lattice-gas model developed to account for these observations is based on the assumption that the sulfate ($SO_4^{2-}$) or bisulfate ($HSO_4^-$) coordinates the Rh(111) surface through three of its oxygen atoms, with the fourth S–O bond pointing away from the surface, as is also the most likely adsorption geometry on Au(111) [8–12,41]. This adsorption geometry gives the sulfate a "footprint" in the shape of an approximately equilateral triangle with an O-O distance of 2.4 Å [45], reasonably matching the lattice constant for the triangular Rh(111) unit cell, 2.69 Å [31]. This high symmetry is somewhat distorted for bisulfate. Based on spectroscopic information, it is likely that the (bi)sulfate coordinates the rhodium atoms directly through three oxygen atoms, thus bringing the sulfur directly above the three-fold hollow sites [46]. Integration of the CV curves indicates that the hydrogen saturation coverage in the negative-potential region corresponds to one elementary charge per Rh(111) unit cell, and that most of the surface hydrogen is desorbed in the same potential range where (bi)sulfate becomes adsorbed. As for platinum [32], we expect that the hydrogen is adsorbed at on-top sites. In the resulting model the (bi)sulfate is modeled as hard hexagons [8–12,47], and the hydrogen as point particles, and finite effective pair interactions out to fourth-nearest neighbors are included to stabilize the $(\sqrt{3}\times\sqrt{7})$ phase. In addition, three-particle interactions [27] are included to discourage the formation of the $(\sqrt{3}\times\sqrt{3})$ phase.

The effective lattice-gas interactions can be estimated as described for urea on Pt(111) in Sec. 2, and simulations were performed on the workstation cluster described in Sec. 2 for systems up to 30×60 triangular-lattice unit cells. For the same reasons of computational speed discussed in Sec. 2, we used a heat-bath algorithm with updates of clusters consisting of two nearest-neighbor sites, which have a total of six possible adsorption states. Preliminary simulation results for $\Theta_S$ in a simplified model, in which the hydrogen and the sulfate are competitively adsorbed on the same sites, are shown in Fig. 6, together with the experimental RCM results. These simulation results agree quite reasonably with the (bi)sulfate saturation coverage, as well as with the general shape of the experimental coverage curves in the transition potential region and with the dependence of the transition potential on the bulk concentration of sulfuric acid. The "jaggedness" of the simulated coverage curves in the 0.17–0.20 ML region is a consequence of the relatively short simulation runs used ($2.4\times10^5$ MCSS per point). Further simulations are in progress. A representative equilibrium configuration for the simplified model in the ordered $(\sqrt{3}\times\sqrt{7})$ phase region is shown in Fig. 7.

## 4  Discussion

In this paper we have discussed two particular applications of lattice-gas models with effective lateral interactions to electrochemical adsorption of small molecules and ions on well-characterized single-crystal surfaces. Detailed experimental information is available for both systems studied. A characteristic aspect of these systems is the good geometric fit between the adsorbent surface lattice and the main adsorbate. This allows the formation of ordered adsorbate phases commensurate with the substrate, which can be observed both by *in situ* atomic-scale microscopies and by *ex situ* scattering techniques. The associated phase transitions give rise to sharp features in experimentally measured adsorption isotherms and voltammograms. The static aspects of these experimental features are well reproduced by our lattice-gas models with effective interaction constants. The interactions in the physical system arise from a number of different sources, including mediation through the substrate electrons, through phonons, and through the fluid near the surface. Their calculation from first principles is therefore still not feasible in general. The alternative route followed here provides a microscopic picture of the adsorbate structure, as well as a procedure for estimating approximate effective interaction energies from experimentally observed thermodynamic quantities. These effective interactions are generally taken as the minimal set necessary to ensure the stability of the observed ordered phases. Although one cannot expect such estimates to be highly accurate, we find that they lead to models with considerable predictive power regarding the dependences of observed thermodynamic quantities on bulk solute concentrations and, to a somewhat lesser degree, on temperature. Our studies further demonstrate the importance of the geometric fit between substrate and adsorbate. This provides a powerful means of predicting from considerations of local bonding geometry which substrate-adsorbate combinations are most likely to give rise to commensurate ordered adsorbate phases. Since our methods are not very computationally intensive, they should be well suited for data analysis by experimental groups.

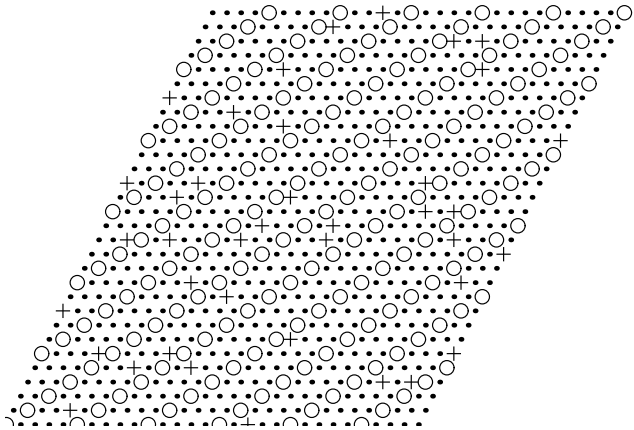

Fig. 7. A typical equilibrium configuration for our simplified model for (bi)sulfate adsorption on Rh(111), generated by MC in the ordered $(\sqrt{3}\times\sqrt{7})$ phase region at $-150\,\mathrm{mV}$. (Bi)sulfate ions are shown as $\bigcirc$, hydrogen as $+$, and empty lattice sites as $\bullet$.

Lattice-gas modeling along the lines described in this paper provides basic information about the structure and stability of ordered adsorbate phases on single-crystal electrode surfaces, that is not readily available by other theoretical or numerical methods at the present time. Such knowledge in turn is necessary to develop techniques to manipulate the adsorption of ordered phases in applied fields, such as electrocatalysis.

## Acknowledgement

We acknowledge useful discussions with L. Blum, K. Franaszczuk, M. Ito, A. Tadjeddine, M.F. Toney, and J.X. Wang, correspondence with K. Itaya, comments on the manuscript by B.M. Gorman, C.C.A. Günther, M.A. Novotny, and S.W. Sides, and technical and artistic assistance with Ref. [19] and Fig. 3 by E. Pepke and D. Poindexter. We thank the authors of Ref. [43] for allowing us to use their data before publication. This research was supported by the Florida State University Supercomputer Computations Research Institute under US Department of Energy Contract No. DE-FC05-85ER25000, by the Florida State University Center for Materials Science and Technology, and by the University of Illinois Frederick Seitz Materials Research Laboratory under US Department of Energy Contract No. DE-AC02-76ER01198. Work at FSU was also supported by US National Science Foundation Grants No. DMR-9013107 and DMR-9315969.